\documentclass{kluwer}    
\usepackage{graphicx}

\begin{document}
\begin{article}
\begin{opening}
\title{Large--scale Structure, Theory and Statistics}
\author{Peter \surname{Coles}}
\runningauthor{Peter Coles} \runningtitle{Large-scale Structure}
\institute{School of Physics \& Astronomy, University of
Nottingham, University Park, Nottingham NG7 2RD, United Kingdom.}
\date{December 25, 2000}

\begin{abstract}
I review the standard paradigm for understanding the formation and
evolution of cosmic structure, based on the gravitational
instability of dark matter, but many variations on this basic
theme are viable. Despite the great progress that has undoubtedly
been made, steps are difficult because of uncertainties in the
cosmological parameters, in the modelling of relevant physical
processes involved in galaxy formation, and perhaps most
fundamentally in the relationship between galaxies and the
underlying distribution of matter. For the foreseeable future,
therefore, this field will be led by observational developments
allowing model parameters to be tuned and, hopefully, particular
scenarios falsified. In these lectures I focus on two ingredients
in this class of models: (i) the role of galaxy bias in
interpreting clustering data; and (ii) the statistical properties
of the initial fluctuations. In the later case, I discuss some
ideas as to how the standard assumption - that the primordial
density fluctuations constitute a Gaussian random field - can be
tested using measurements galaxy clustering and the cosmic
microwave background.
\end{abstract}

\end{opening}

\section{Introduction}

The local Universe displays a rich hierarchical pattern of galaxy
clusters and superclusters~\cite{lasc}. The early Universe,
however, was almost smooth, with only slight ripples seen in the
cosmic microwave background radiation~\cite{Cobe}. Models of the
evolution of structure link these observations through the effect
of gravity, because the small initially overdense fluctuations
attract additional mass as the Universe expands~\cite{peeb}.
During the early stages, the ripples evolve independently, like
linear waves on the surface of deep water. As the structures grow
in mass, they interact with other in non-linear ways, more like
nonlinear waves breaking in shallow water.

The expansion of the Universe renders the cosmological version of
gravitational instability very slow, a power-law in time rather
the exponential growth that develops in a static background. This
slow rate has the important consequence that the evolved
distribution of mass still retains significant memory of the
initial state. This, in turn, has two consequences for theories of
structure formation. One is that a detailed model must entail a
complete prescription for the form of the initial conditions, and
the other is that observations made at the present epoch allow us
to probe the primordial fluctuations and thus test the theory.

Cosmology is now poised on the threshold of a data explosion
which, if harnessed correctly, should yield a definitive answer to
the question of initial fluctuations. The next generation of
galaxy survey projects will furnish data sets capable answering
many of the outstanding issues in this field including that of the
form of the initial fluctuations.  Planned CMB missions, including
the Planck Surveyor, will yield higher-resolution maps of the
temperature anisotropy pattern that will subject cosmological
models to still more detailed scrutiny.

In these lectures I discuss the formation of large-scale structure
from a general point of view, but emphasizing two of the most
important gaps in our current knowledge and suggesting how these
might be answered if the new data can be exploited efficiently. I
begin with a general review of the theory in Section 2, discuss
(briefly) possible observational developments in Section 3.
Section 4 addresses the form and statistics of primordial density
perturbations, particularly the question whether they are
gaussian. In Section 5 I discuss uncertainties in the relationship
between the distribution of galaxies and that of mass and some
recent developments in the understanding of that relationship in a
statistical sense.

\section{Cosmological Structure Formation}

\subsection{Basics of the Big Bang}

The Big Bang theory is built upon the Cosmological Principle, a
symmetry principle that requires the Universe on large scales to
be both homogeneous and isotropic. Space-times consistent with
this requirement can be described by the Robertson--Walker metric
\begin{equation}
{\rm d}s_{\rm FRW}^2 = c^2 {\rm d}t^2 - a^2(t)\left({{\rm
d}r^2\over 1 - \kappa r^2} + r^2 {\rm d}\theta^2 + r^2\sin^2\theta
{\rm d}\phi^2\right)  , \label{eq:l1a}
\end{equation}
where $\kappa$ is the spatial curvature, scaled so as to take the
values $0$ or $\pm 1$. The case $\kappa=0$ represents  flat space
sections, and the other two cases are  space sections of constant
positive or negative curvature, respectively. The time coordinate
$t$ is called {\em cosmological proper time} and it is singled out
as a preferred time coordinate by the property of spatial
homogeneity. The quantity $a(t)$,  the {\em cosmic scale factor},
describes the overall expansion of the universe  as a function of
time. If light emitted at time $t_{\rm e}$ is received by an
observer at $t_0$ then the redshift $z$ of the source is given by
\begin{equation}
1+z = \frac{a(t_0)}{a(t_{\rm e})}.
\end{equation}

The dynamics of an FRW universe are determined by the Einstein
gravitational field equations which become
\begin{eqnarray}
3\left( \frac{\dot{a}}{a} \right)^{2} & = & 8\pi G\rho - {3\kappa
c^{2} \over a^2} + \Lambda,\\ {\ddot{a}\over a} & = & - {4\pi
G\over 3} \left(\rho + 3 \frac{p}{c^2}\right) + {\Lambda\over 3},
\\ \dot{\rho}& =& - 3 {\dot{a}\over a}\left(\rho + \frac{p}{c^2}
\right). \label{eq:l1b}
\end{eqnarray}
These equations  determine the time evolution of the cosmic scale
factor $a(t)$ (the dots denote derivatives with respect to
cosmological proper time $t$) and therefore describe the global
expansion or contraction of the universe. The behaviour of these
models can further be parametrised in terms of the Hubble
parameter $H=\dot{a}/a$ and the density parameter $\Omega=8\pi
G\rho/3H^2$, a suffix $0$ representing the value of these
quantities at the present epoch when $t=t_0$.

\subsection{Linear Perturbation Theory}

In order to understand how structures form we need to consider the
difficult problem of dealing with the evolution of inhomogeneities
in the expanding Universe. We are helped in this task by the fact
that we expect such inhomogeneities to be of very small amplitude
early on so we can adopt a kind of perturbative approach, at least
for the early stages of the problem. If the length scale of the
perturbations is smaller than the effective cosmological horizon
$d_H=c/H_0$, a Newtonian treatment of the subject  is expected to
be valid.  If the mean free path of a particle is small, matter
can be treated as an ideal fluid and the Newtonian equations
governing the motion of gravitating particles in an expanding
universe can be written in terms of ${\bf x} = {\bf r} / a(t)$
(the comoving spatial coordinate, which is fixed for observers
moving with the Hubble expansion), ${\bf v} = \dot {{\bf r}} - H
{\bf r} = a\dot {{\bf x}}$ (the peculiar velocity field,
representing departures of the matter motion from pure Hubble
expansion), $\phi ({\bf x} , t)$ (the peculiar Newtonian
gravitational potential, i.e. the fluctuations in potential with
respect to the homogeneous background) and $\rho ({\bf x}, t)$
(the matter density). Using these variables
 we obtain, first, {\em the Euler equation}:
\begin{equation}
{\partial (a{\bf v})\over \partial t} + ({\bf v}\cdot{\bf
\nabla_x}){\bf v} = - {1\over \rho}{\bf \nabla_x} p - {\bf
\nabla_x}\phi~. \label{eq:Euler}
\end{equation}
The second term on the right-hand side of equation
(\ref{eq:Euler}) is the peculiar gravitational force, which can be
written in terms of ${\bf g} = -{\bf \nabla_x}\phi/a$, the
peculiar gravitational acceleration of the fluid element. If the
velocity flow is irrotational, ${\bf v}$ can be rewritten in terms
of a velocity potential $\phi_v$: ${\bf v} = - {\bf \nabla_x}
\phi_v/a$. Next we have the {\em continuity equation}:
\begin{equation}
{\partial\rho\over \partial t} + 3H\rho + {1\over a} {\bf
\nabla_x} (\rho{\bf v}) = 0, \label{eq:continuity}
\end{equation}
which expresses the conservation of matter, and finally the {\em
Poisson equation}:
\begin{equation}
{\bf \nabla_x}^2\phi = 4\pi G a^2(\rho - \rho_0) = 4\pi
Ga^2\rho_0\delta, \label{eq:Poisson}
\end{equation}
describing Newtonian gravity.  Here $\rho_0$ is the mean
background density, and
\begin{equation}
\delta \equiv \frac{\rho-\rho_0}{\rho_0}
\end{equation}
is the {\em density contrast}.

The next step is  to linearise the Euler, continuity and Poisson
equations by perturbing physical quantities defined as functions
of Eulerian coordinates, i.e. relative to an unperturbed
coordinate system. Expanding $\rho$, ${\bf v}$ and $\phi$
perturbatively and keeping only the first-order terms in equations
(\ref{eq:Euler}) and (\ref{eq:continuity}) gives the linearised
continuity equation:
\begin{equation}
{\partial\delta\over \partial t} = - {1\over a}{\bf \nabla_x}\cdot
{\bf v},
\end{equation}
which can be inverted, with a suitable choice of boundary
conditions, to yield
\begin{equation}
\delta = - {1\over a H f}\left({\bf \nabla_x}\cdot{\bf v}\right).
\label{eq:l9}
\end{equation}
The function $f\simeq \Omega_0^{0.6}$; this is simply a fitting
formula to the full solution \cite{peeb}. The linearised Euler and
Poisson equations are
\begin{equation}
{\partial {\bf v}\over\partial t} + {\dot a\over a}{\bf v} = -
{1\over \rho a}{\bf \nabla_x} p -{1\over a}{\bf \nabla_x}\phi,
\label{eq:l10}
\end{equation}
\begin{equation}
{\bf \nabla_x}^2\phi = 4\pi G a^2\rho_0\delta; \label{eq:l11}
\end{equation}
$|{\bf v}|, |\phi|, |\delta| \ll 1$ in equations (\ref{eq:l9}),
(\ref{eq:l10}) \& (\ref{eq:l11}). From these equations, and if one
ignores pressure forces, it is easy to obtain an equation for the
evolution of $\delta$:
\begin{equation}
\ddot\delta + 2H\dot\delta - {3\over 2}\Omega H^2\delta = 0.
\label{eq:l13b}
\end{equation}
For a spatially flat universe dominated by pressureless matter,
$\rho_0(t) = 1/6\pi Gt^2$ and equation (\ref{eq:l13b}) admits two
linearly independent power law solutions $\delta({\bf x},t) =
D_{\pm}(t)\delta({\bf x})$, where $D_+(t) \propto a(t) \propto
t^{2/3}$  is the growing  mode and $D_-(t) \propto t^{-1}$ is the
decaying mode.

\subsection{Primordial density fluctuations}
The above considerations apply to the evolution of a single
Fourier mode of the density field $\delta({\bf x}, t) =
D_+(t)\delta({\bf x})$. What is more likely to be relevant,
however, is the case of a superposition of waves, resulting from
some kind of stochastic process in which he density field consists
of a  superposition of such modes with different amplitudes. A
statistical description of the initial perturbations is therefore
required, and any comparison between theory and observations will
also have to be statistical.

The spatial Fourier transform of $\delta({\bf x})$ is
\begin{equation}
\tilde{\delta}({\bf k}) = \frac{1}{(2\pi)^{3}} \int {\rm d}^{3}
{\bf x}e^{- i{\bf k}\cdot{\bf x}} \delta({\bf x}). \label{eq:lad1}
\end{equation}
It is useful to specify the properties of $\delta$ in terms of
$\tilde{\delta}$. We can define the {\em power-spectrum} of the
field to be (essentially) the variance of the amplitudes at a
given value of ${\bf k}$:
\begin{equation}
\langle \tilde{\delta}({\bf k}_1) \tilde{\delta}({\bf k}_2)
\rangle = P(k_1) \delta^{D} ({\bf k}_1+{\bf k}_2), \label{eq:lad4}
\end{equation}
where $\delta^{D}$ is the Dirac delta function; this rather
cumbersome definition takes account of the translation symmetry
and reality requirements for $P(k)$; isotropy is expressed by
$P({\bf k})=P(k)$. The analogous quantity in real space is called
the two-point correlation function or, more correctly, the
autocovariance function, of $\delta({\bf x})$:
\begin{equation}
\langle \delta({\bf x}_1) \delta({\bf x}_2) \rangle =\xi (|{\bf
x}_1-{\bf x_2}|) = \xi({\bf r})=\xi(r), \label{eq:lad5}
\end{equation}
which is itself related to the power spectrum via a Fourier
transform. The shape of the initial fluctuation spectrum, is
assumed to be imprinted on the universe at some arbitrarily early
time. Many versions of the inflationary scenario for the very
early universe \cite{guth,gp} produce  a power-law form
\begin{equation}
P(k)=Ak^{n}, \label{eq:lad10}
\end{equation}
with a preference in some cases for the Harrison--Zel'dovich form
with $n=1$ \cite{harrison,zeld}. Even if inflation is not the
origin of density fluctuations, the form (\ref{eq:lad10}) is a
useful phenomenological model for the fluctuation spectrum. These
considerations specify the shape of the fluctuation spectrum, but
not its amplitude. The discovery of temperature fluctuations in
the CMB \cite{Cobe} has plugged that gap.

The power-spectrum is particularly important because it provides a
complete statistical characterisation of a particular kind of
stochastic process: a {\em Gaussian random field}. This class of
field is the generic prediction of inflationary models, in which
the density perturbations are generated by Gaussian quantum
fluctuations  in a scalar field during the inflationary epoch
\cite{gp,brand}.

\subsection{The transfer function}
We have hitherto assumed that the effects of pressure and other
astrophysical processes on the gravitational evolution of
perturbations are negligible. In fact, depending on the form of
any dark matter, and the parameters of the background cosmology,
the growth of perturbations on particular length scales can be
suppressed relative to the growth laws discussed above.

We need first to specify the fluctuation mode. In cosmology, the
two relevant alternatives are {\em adiabatic} and {\em
isocurvature}. The former involve coupled fluctuations in the
matter and radiation component in such a way that the entropy does
not vary spatially; the latter have zero net fluctuation in the
energy density and involve entropy fluctuations. Adiabatic
fluctuations are the generic prediction from inflation and form
the basis of most currently fashionable models, although
interesting work has been done recently on isocurvature models
\cite{peeb1, peeb2}.

In the classical Jeans instability, pressure inhibits the growth
of structure on scales smaller than the distance traversed by an
acoustic wave during the free-fall collapse time of a
perturbation. If there are collisionless particles of hot dark
matter, they can travel rapidly through the background and this
free streaming can damp away perturbations completely. Radiation
and relativistic particles may also cause kinematic suppression of
growth. The imperfect coupling of photons and baryons can also
cause dissipation of perturbations in the baryonic component. The
net effect of these processes, for the case of statistically
homogeneous initial Gaussian fluctuations, is to change the shape
of the original power-spectrum in a manner described by a simple
function of wave-number -- the transfer function $T(k)$ -- which
relates the processed power-spectrum $P(k)$ to its primordial form
$P_0(k)$ via $P(k) = P_0(k)\times T^2(k)$. The results of full
numerical calculations of all the physical processes we have
discussed can be encoded in the transfer function of a particular
model \cite{bbks}. For example, fast moving or `hot' dark matter
particles (HDM) erase structure on small scales by the
free-streaming effects mentioned above so that $T(k)\rightarrow 0$
exponentially for large $k$; slow moving or `cold' dark matter
(CDM) does not suffer such strong dissipation, but there is a
kinematic suppression of growth on small scales (to be more
precise, on scales less than the horizon size at matter--radiation
equality); significant small-scale power nevertheless survives in
the latter case. These two alternatives thus furnish two very
different scenarios for the late stages of structure formation:
the `top--down' picture exemplified by HDM first produces
superclusters, which subsequently fragment to form galaxies; CDM
is a `bottom--up' model because small-scale structures form first
and then merge to form larger ones. The general picture that
emerges is that, while the amplitude of each Fourier mode remains
small, i.e. $\delta({\bf k})\ll 1$, linear theory applies. In this
regime, each Fourier mode evolves independently and the
power-spectrum therefore just scales as
\begin{equation}
P(k,t)=P(k,t_1) {D_{+}^{2}(k,t)\over D_{+}^2(k,t_1)} = P_0(k)
T^{2} (k) {D_{+}^{2}(k,t)\over D_{+}^2(k,t_1)}~.
\end{equation}
For scales larger than the Jeans length, this means that the shape
of the power-spectrum is preserved during linear evolution.

\subsection{Beyond linear theory}
The linearised equations of motion  provide an excellent
description of gravitational instability at very early times when
density fluctuations are still small ($\delta \ll 1$). The linear
regime of gravitational instability breaks down when $\delta$
becomes comparable to unity, marking the commencement of the {\it
quasi-linear} (or weakly non-linear) regime. During this regime
the density contrast may remain small ($\delta < 1$), but the
phases of the Fourier components $\delta_{\bf k}$ become
substantially different from their initial values resulting in the
gradual development of a non-Gaussian distribution function if the
primordial density field was Gaussian. In this regime the shape of
the power-spectrum changes by virtue of a complicated cross-talk
between different wave-modes. Analytic methods are available for
this kind of problem \cite{sc95}, but the usual approach is to use
$N$-body experiments for strongly non-linear analyses
\cite{defw,virgo}.

Further into the non-linear regime, bound structures form. The
baryonic content of these objects may then become important
dynamically: hydrodynamical effects (e.g. shocks), star formation
and heating and cooling of gas all come into play. The spatial
distribution of galaxies may therefore be very different from the
distribution of the (dark) matter, even on large scales. Attempts
are only just being made to model some of these processes with
cosmological hydrodynamics codes \cite{hydro}, but it is some
measure of the difficulty of understanding the formation of
galaxies and clusters that most studies have only just begun to
attempt to include modelling the detailed physics of galaxy
formation. In the front rank of theoretical efforts in this area
are the so-called semi-analytical models which encode simple rules
for the formation of stars within a framework of merger trees that
allows the hierarchical nature of gravitational instability to be
explicitly taken into account \cite{anal}.

The usual approach is instead simply to assume that the point-like
distribution of galaxies, galaxy clusters or whatever,
\begin{equation}
n({\bf r})=\sum_i\delta_D({\bf r} - {\bf r}_i),
\end{equation}
bears a simple functional relationship to the underlying
$\delta({\bf r})$.
 An assumption often invoked is that relative fluctuations in the object number
counts and matter density fluctuations are proportional to each
other, at least within sufficiently large volumes, according to
the {\em linear biasing} prescription:
\begin{equation}
{\delta n({\bf r})\over \bar n}~=~b\,{\delta \rho({\bf r})\over
\bar \rho}\,, \label{eq:linb}
\end{equation}
where $b$ is what is usually called the biasing parameter.
Alternatives, which are not equivalent, include the high-peak
model (\cite{k84,bbks}) and the various local bias models
\cite{lbias}. Non-local biases are possible, but it is rather
harder to construct such models \cite{coop}.  If one is prepared
to accept an {\it ansatz} of the form (\ref{eq:linb}) then one can
use linear theory on large scales to relate galaxy clustering
statistics to those of the density fluctuations, e.g.
\begin{equation}
P_{\rm gal}(k)=b^{2}P(k).
\end{equation}
This approach is the one most frequently adopted in practice, but
the community is becoming increasingly aware of its severe
limitations. A simple parametrisation of this kind simply cannot
hope to describe realistically the relationship between galaxy
formation and environment \cite{stoch}. I will return to this
question in Section 5.

\subsection{Models of structure formation}

It should now be clear that models of structure formation involve
many ingredients which interact in a complicated way. In the
following list, notice that most of these ingredients involve at
least one assumption that may well turn out not to be true:
\begin{enumerate}
\renewcommand{\theenumi}{(\arabic{enumi})}
\setcounter{enumi}{0}
\item A background cosmology. This basically means a choice
of $\Omega_0$, $H_0$ and $\Lambda$, assuming we are prepared to
stick with the Robertson--Walker metric (1) and the Einstein
equations  (3)-(5).
\item An initial fluctuation spectrum. This is usually
taken to be a power-law, but may not be. The most common choice is
$n=1$.
\item A choice of fluctuation mode: usually adiabatic.
\item A statistical distribution of fluctuations. This is
often assumed to be Gaussian.
\item The transfer function, which requires knowledge of the
relevant proportions of `hot', `cold' and baryonic material as
well as the number of relativistic particle species.
\item A `machine' for handling non-linear evolution,
so that the distribution of galaxies and other structures can be
predicted. This could be an $N$-body or hydrodynamical code, an
approximated dynamical calculation or simply, with fingers
crossed, linear theory.
\item A prescription for relating fluctuations in mass to
fluctuations in light, frequently the linear bias model.
\end{enumerate}

Historically speaking, the first  model incorporating non-baryonic
dark matter to be seriously considered was the hot dark matter
({\bf HDM}) scenario, in which the universe is dominated by a
massive neutrino with mass around 10--30 eV. This scenario has
fallen into disrepute because the copious free streaming it
produces smooths the matter fluctuations on small scales and means
that galaxies form very late. The favoured alternative for most of
the 1980s was the cold dark matter ({\bf CDM}) model in which the
dark matter particles undergo negligible free streaming owing to
their higher mass or non-thermal behaviour. A `standard' CDM model
({\bf SCDM}) then emerged in which the cosmological parameters
were fixed at $\Omega_0=1$ and $h=0.5$, the spectrum was of the
Harrison--Zel'dovich form with $n=1$ and a significant bias,
$b=1.5$ to $2.5$, was required to fit the observations
\cite{defw}.

The SCDM model was ruled out by a combination of the COBE-inferred
amplitude of primordial density fluctuations, galaxy clustering
power-spectrum estimates on large scales, cluster abundances and
small-scale velocity dispersions \cite{pd96}. It seems the
standard version of this theory simply has a transfer function
with the wrong shape to accommodate all the available data with an
$n=1$ initial spectrum. Nevertheless, because CDM is such a
successful first approximation and seems to have gone a long way
to providing an answer to the puzzle of structure formation, the
response of the community has not been to abandon it entirely, but
to seek ways of relaxing the constituent assumptions in order to
get a better agreement with observations. Various possibilities
have been suggested.

If the total density is reduced to $\Omega_0\simeq 0.3$, which is
favoured by many  arguments, then the size of the horizon at
matter--radiation equivalence increases compared with SCDM and
much more large-scale clustering is generated. . This is called
the open cold dark matter model, or {\bf OCDM} for short. Those
unwilling to dispense with the inflationary predeliction for flat
spatial sections have invoked $\Omega_0=0.2$ and a positive
cosmological constant \cite{lcdm} to ensure that $k=0$; this can
be called {\bf $\Lambda$CDM} and is apparently also favoured by
observations of distant supernovae \cite{hzsn}. Much the same
effect on the power spectrum may also be obtained in $\Omega=1$
CDM models if matter-radiation equivalence is delayed, such as by
the addition of an additional relativistic particle species. The
resulting models are usually called {\bf $\tau$CDM} \cite{taucdm}.

Another alternative to SCDM  involves a mixture of hot and cold
dark matter ({\bf CHDM}), having perhaps $\Omega_{\rm hot}=0.3$
for the fractional density contributed by the hot particles. For a
fixed large-scale normalisation, adding a hot component has the
effect of suppressing the power-spectrum amplitude at small
wavelengths \cite{chdmr}. A variation on this theme would be to
invoke a `volatile' rather than `hot' component of matter produced
by the decay of a heavier particle \cite{cvdmr}. The non-thermal
character of the decay products results in subtle differences in
the shape of the transfer function in the ({\bf CVDM}) model
compared to the {\bf CHDM} version. Another possibility is to
invoke non-flat initial fluctuation spectra, while keeping
everything else in SCDM fixed. The resulting `tilted' models, {\bf
TCDM}, usually have $n<1$ power-law spectra for extra large-scale
power and, perhaps, a significant fraction of tensor perturbations
\cite{lc92}. Models have also been constructed in which
non-power-law behaviour is invoked to produce the required extra
power: these are the broken scale-invariance ({\bf BSI}) models
\cite{bsir}.

But diverse though this collection of alternative models may seem,
it does not include models where the assumption of Gaussian
statistics is relaxed. This is at least as important as the other
ingredients which have been varied in some of the above models.
The reason for this is that fully-specified non-Gaussian models
are hard to construct, even if they are based on purely
phenomenological considerations \cite{ng1,ng2}. Models based on
topological defects rather than inflation generally produce
non-Gaussian features  but are computationally challenging
\cite{ng3}. A notable exception to this dearth of alternatives is
the ingenious isocurvature model of Peebles \cite{peeb1,peeb2}.

\begin{figure}
  \begin{center}
    \caption{Some of the candidate models described in the text, as simulated by the Virgo
    consortium. Notice that SCDM shows very different structure at $z=0$ than the three alternatives
    shown. The models also differ significantly at different epochs. These simulations show the distribution
    of dark matter only.}
    \label{fig:fig1}
  \end{center}
\end{figure}

\section{Observational Prospects}

\subsection{Redshift surveys}
In 1986, the CfA survey \cite{cfa}  was the `state-of-the-art',
but this contained redshifts of only around 2000 galaxies with a
maximum recession velocity of $15~000$ km s$^{-1}$. The Las
Campanas survey contains around six times as many galaxies, and
goes out to a velocity of $60~000$ km s$^{-1}$ \cite{lasc}. At
present, redshifts of around $10^{5}$ galaxies are available. The
next generation of redshift surveys, prominent among which are the
Sloan Digital Sky Survey \cite{sloan} of about one million galaxy
redshifts and an Anglo-Australian collaboration using the
two-degree field (2DF) \cite{2df}; these surveys exploit
multi-fibre methods which can obtain 400 galaxy spectra in one go,
and will increase the number of redshifts by about two orders of
magnitude over what is currently available.

Quantitative measures of spatial clustering obtained from these
data sets offer the simplest method of probing $P(k)$, assuming
that these objects are related in some well-defined way to the
mass distribution and this, through the transfer function, is one
way of constraining cosmological parameters.

\subsection{The Galaxy Power-spectrum}
Although the traditional tool for studying galaxy clustering is
the two-point correlation function, $\xi(r)$ \cite{peeb}, defined
by \begin{equation} dP_{12}=n^2[1+\xi(r)]dV_1 dV_2, \end{equation}
the (small) joint probability of finding two galaxies in the
(small) volumes $dV_1$ and $dV_2$ separated by a distance $r$ when
the mean number-density of galaxies is $n$. Most modern analyses
concentrate instead upon its Fourier transform, the power-spectrum
$P(k)$. This is especially useful because it is the power-spectrum
which is predicted directly in cosmogonical models incorporating
inflation and dark matter. For example, Peacock \& Dodds have
recently made compilations of power-spectra of different kinds of
galaxy and cluster redshift samples and, for comparison, a
deprojection of the APM $w(\theta)$ \cite{pd96}. Within the
(considerable) observational errors, and the uncertainty
introduced by modelling of the bias, all the data lie roughly on
the same curve. A consistent picture thus seems to have emerged in
which galaxy clustering extends over larger scales than is
expected in the standard CDM scenario. Considerable uncertainty
nevertheless remains about the shape of the power spectrum on very
large scales.

\subsection{The abundances of objects}
In addition to their spatial distribution, the number-densities of
various classes of cosmic objects as a function of redshift can be
used to constrain the shape of the power-spectrum. In particular,
if objects are forming by hierarchical merging there should be
fewer objects of a given mass at high $z$ and  more objects with
lower mass. This can be made quantitative fairly simply, using an
analytic method \cite{ps74}. Although this kind of argument can be
applied to many classes of object \cite{ma97}, it potentially
yields the strongest constraints when applied to galaxy clusters.
At the moment, results are controversial, but the evolution of
cluster numbers with redshift is such  sensitive probe of
$\Omega_0$ so that future studies of high-redshift clusters may
yield more definitive results \cite{ecf96,blanch}.

\subsection{High-redshift clustering}

It is evident from Figure 1 that, although the three non-SCDM
models are similar at $z=0$, differences between them are marked
at higher redshift. This suggests the possibility of using
measurements of galaxy clustering at high redshift to distinguish
between models and reality. This has now become possible, with
surveys of galaxies at $z \sim 3$ already being constructed
\cite{steidel}. Unfortunately, the interpretation of these new
data is less straightforward than one might have imagined. If the
galaxy distribution is biased at $z=0$ then the bias is expected
to grow with $z$ \cite{defw}. If galaxies are rare peaks now, they
should have been even rarer at high $z$. There are also many
distinct possibilities as to how the bias might evolve with
redshift \cite{mosc}. Theoretical uncertainties therefore make it
difficult to place stringent constrains on models, although with
more data and better theoretical modelling, high-redshift
clustering measurements will play a very important role in
forthcoming years.

\subsection{Higher-order Statistics}

The galaxy power-spectrum has rightly played a central role in the
development of this subject, but the information it contains is in
fact rather limited. In more precise terms, it is called a
second-order statistic as it contains information equivalent to
the second moment (i.e. variance) of a random variable.
Higher-order statistics would be necessary to provide a complete
statistical description of clustering pattern and these generally
require large, well-sampled data sets \cite{sc95}. One
particularly promising set of descriptors emerge from the
realisation \cite{skew1,skew2,skew3} that higher-order moments
grow by gravitational instability in a manner that couples
directly to the growth of the variance. This offers the prospect
of being able to distinguish between genuine clustering produced
by gravity and clustering induced by bias.

\subsection{Peculiar Motions}
There are various ways in which it is possible to use information
about the velocities of galaxies to constrain models \cite{sw95}.
Probably the most useful information pertains to large-scale
motions, as small-scale data populate the highly nonlinear regime.

The basic principle is that velocities are induced by fluctuations
in the total mass, not just the galaxies. Comparing measured
velocities with measured fluctuations in galaxies with measured
fluctuations in galaxy counts, it is possible to constrain both
$\Omega$ and $b$. From equations (\ref{eq:l9})--(\ref{eq:l11}) it
emerges that
\begin{equation}
{\bf v} = - {2f\over 3\Omega H a}{\bf \nabla_x}\phi + {\mbox{
const}\over a(t)}, \label{eq:l15}
\end{equation}
which demonstrates that the velocity flow associated with the
growing mode in the linear regime is curl-free, as it can be
expressed as the gradient of a scalar potential function. Notice
also that the induced velocity depends on $\Omega$. This is the
basis of a method for estimating $\Omega$ which is known as POTENT
\cite{potent}. Since all matter gravitates, not just the luminous
material, there is a hope that methods such as this can break the
degeneracy between clustering induced by gravity and that induced
statistically, by bias.

These methods are prone to error if there are errors in the
velocity estimates. Perhaps a more robust approach is to use
peculiar motion information indirectly, by the effect they have on
the distribution of galaxies seen in redshift-space (i.e. assuming
total velocity is proportional to distance). The information
gained this way is statistical,  but less prone to systematic
error \cite{zdis}.

\subsection{Gravitational Lensing}

Another class of observations that can help break the degeneracy
between models involves gravitational lensing. The most
spectacular forms of lensing are those producing multiple images
or strong distortions in the form of arcs. These require very
large concentrations of mass and are therefore not so useful for
mapping the structure on large scales. However, there are lensing
effects that are much weaker than the formation of multiple
images. In particular,  distortions producing a shearing of galaxy
images promise much in this regard  \cite{weakl1}. With the advent
of new large CCD detectors, this should soon be realised
\cite{weakl2}, although present constraints are quiet weak.

\subsection{The Cosmic Microwave Background}

I have so far avoided discussion of the cosmic microwave
background, but this probably holds the key to unlocking many of
the present difficulties in large-scale structure models. Although
the COBE data \cite{Cobe} do not constrain the shape of the matter
power spectrum on scales of direct relevance to structures we can
see in the galaxy distribution, finer-scale maps will do so in the
near future. ESA's Planck Explorer and NASA's MAP experiment will
 measure the properties of matter fluctuations in the linear
 without having to worry about the confusion caused by
 non-linearity and bias when galaxy counts are used. It is hoped
 that measurements of particular features in the angular power-spectrum
 of the fluctuations \cite{cmbr} measured  by these experiments will
 pin down the densities of CDM, HDM,  baryons and vacuum energy (i.e. $\Lambda$)
as well as fixing  $\Omega$ and $H$. Experiments such as BOOMERANG
and MAXIMA are already leading to interesting results, but these
are discussed elsewhere in this volume.

\section{Testing Cosmological Gaussianity}

Largely motivated by the idea that they were generated by quantum
fluctuations during a period of inflation, most fashionable models
of structure formation involve the assumption that the initial
fluctuations constitute a Gaussian random field. Mathematically,
this assumption means that all finite- dimensional joint
probability distributions of the density at different spatial
locations can be expressed as multivariate normal distributions.
This is much stronger than the assertion that the distribution of
densities should be a normal distribution. It is quite possible
for a field to have a Gaussian one-point probability distribution
but be non-Gaussian in the sense used here. Testing this form of
multivariate normality in an arbitrary number of dimensions is a
decidedly non-trivial task, but is necessary given the importance
of the assumption. If it can be shown that the large-scale
structure of the Universe is inconsistent with Gaussian initial
data this will have profound implications for fundamental physics.
This issue does not therefore represent a mere exercise in
statistics, but a vital step towards a physical understanding of
the origin and evolution of the large-scale structure of the
Universe.

As well as being physically motivated, the Gaussian assumption has
great advantage that it is a mathematically complete prescription
for all the statistical properties of the initial density field,
once the fluctuation amplitude is specified as a function of scale
through the power-spectrum $P(k)$. In Fourier terms, a Gaussian
random field consists of a stochastic superposition of plane
waves.  The amplitude of each mode, $A_k$, is drawn from a
distribution specified by the power-spectrum and its phase,
$\phi_k$, is uniformly random and independent of the phases of all
other modes. As the fluctuations evolve in time, the density
distribution becomes non-Gaussian. But this departure from
non-Gaussianity depends on gravity being able to move material
from its primordial position. On scales much larger than the
typical scale of such motions, the distribution remains Gaussian.
The distribution of matter today should therefore be highly
non-Gaussian on small scales, gradually tending closer to Gaussian
on progressively larger scales.  Any non-Gaussianity detected at
the present epoch could therefore either be primordial, or
produced dynamically, or could could be imposed by variations in
mass-to-light ratio ("bias"), or all of these. Galaxy clustering
statistics therefore need to be devised that can separate these
different signatures.

The distribution of temperature fluctuations in the cosmic
microwave background (CMB), which was imprinted before significant
gravitational evolution took place, should also retain the
character of the initial statistics. Any non-Gaussianity detected
here could either be primordial, produced by errors in foreground
subtraction or other systematics. Again, tests capable of
distinguishing between these possibilities are required.

Gaussian models have generally fared much better in comparison
with data than others with non-Gaussian initial data, such as
those based on topological defects, although predictions in the
second category of models are harder to come by because of the
much greater calculational difficulties involved. It is fair to
say, however, that as far as existing data are concerned the
large-scale distribution of mass certainly seems to be consistent
with Gaussian statistics. Initially, it also appeared that the
COBE fluctuations in temperature of the CMB were also consistent
with Gaussian primordial perturbations. On the other hand, the
statistical descriptors necessary to carry out a powerful test
against the Gaussian require much higher quality data than has so
far been furnished by galaxy surveys. Moreover, the
non-Gaussianity induced by gravitational evolution, redshift-space
effects, and variations in mass-to-light ratio has complicated the
interpretation of the data, although recent theoretical
developments discussed below should ameliorate these problems.

In the following I discuss a method of quantifying phase
information \cite{cc} and suggest how this information may be
exploited to build novel statistical descriptors that can be used
to mine the sky more effectively than with standard methods.

\subsection{Fourier Description of Cosmological Density Fields} In
most popular versions of the ``gravitational instability'' model
for the origin of cosmic structure, particularly those involving
cosmic inflation~\cite{gp}, the initial fluctuations that seeded
the structure formation process form a Gaussian random
field~\cite{bbks}. Because the initial perturbations evolve
linearly, it is useful to expand $\delta({\bf x})$  as a Fourier
superposition of plane waves:
\begin{equation}
\delta ({\bf x}) =\sum \tilde{\delta}({\bf k}) \exp(i{\bf k}\cdot
{\bf x}).
\end{equation}
The Fourier transform $\tilde{\delta}({\bf k})$ is complex and
therefore possesses both amplitude $|\tilde{\delta} ({\bf k})|$
and phase $\phi_{\bf k}$ where
\begin{equation}
\tilde{\delta}({\bf k})=|\tilde{\delta} ({\bf
k})|\exp(i\phi_{\bf_k}). \label{eq:fourierex}
\end{equation}
Gaussian random fields possess Fourier modes whose real and
imaginary parts are independently distributed. In other words,
they have phase angles $\phi_k$ that are independently distributed
and uniformly random on the interval $[0,2\pi]$. When fluctuations
are small, i.e. during the linear regime, the Fourier modes evolve
independently and their phases remain random. In the later stages
of evolution, however, wave modes begin to couple
together~\cite{peeb}. In this  regime the phases become non-random
and the density field becomes highly non--Gaussian. Phase coupling
is therefore a key consequence of nonlinear gravitational
processes if the initial conditions are Gaussian and a potentially
powerful signature to exploit in statistical tests of this class
of models.

A graphic demonstration of the importance of phases in patterns
generally is given in Fig 2.
\begin{figure}
\centering
\includegraphics[width=0.45\textwidth]{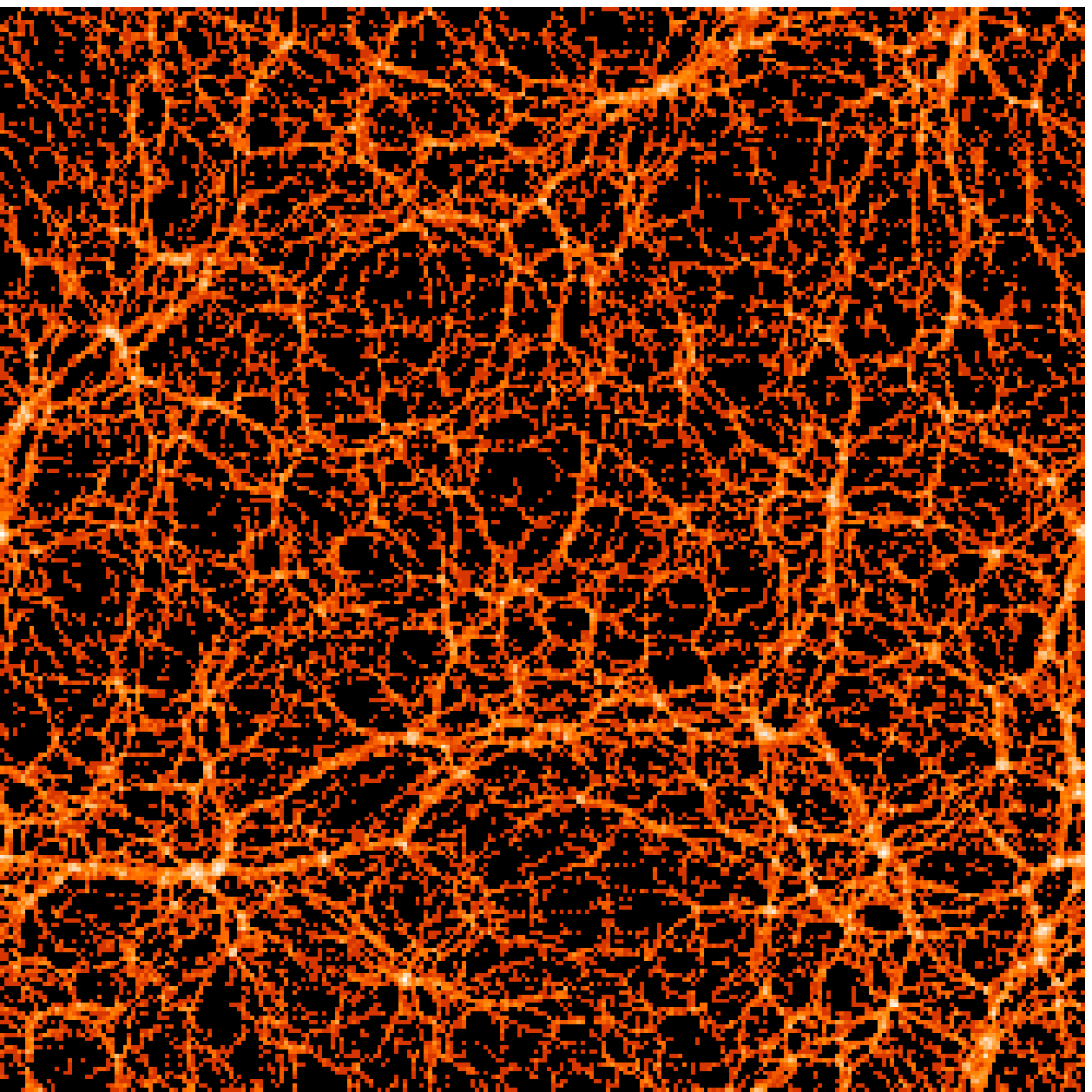}
\includegraphics[width=0.45\textwidth]{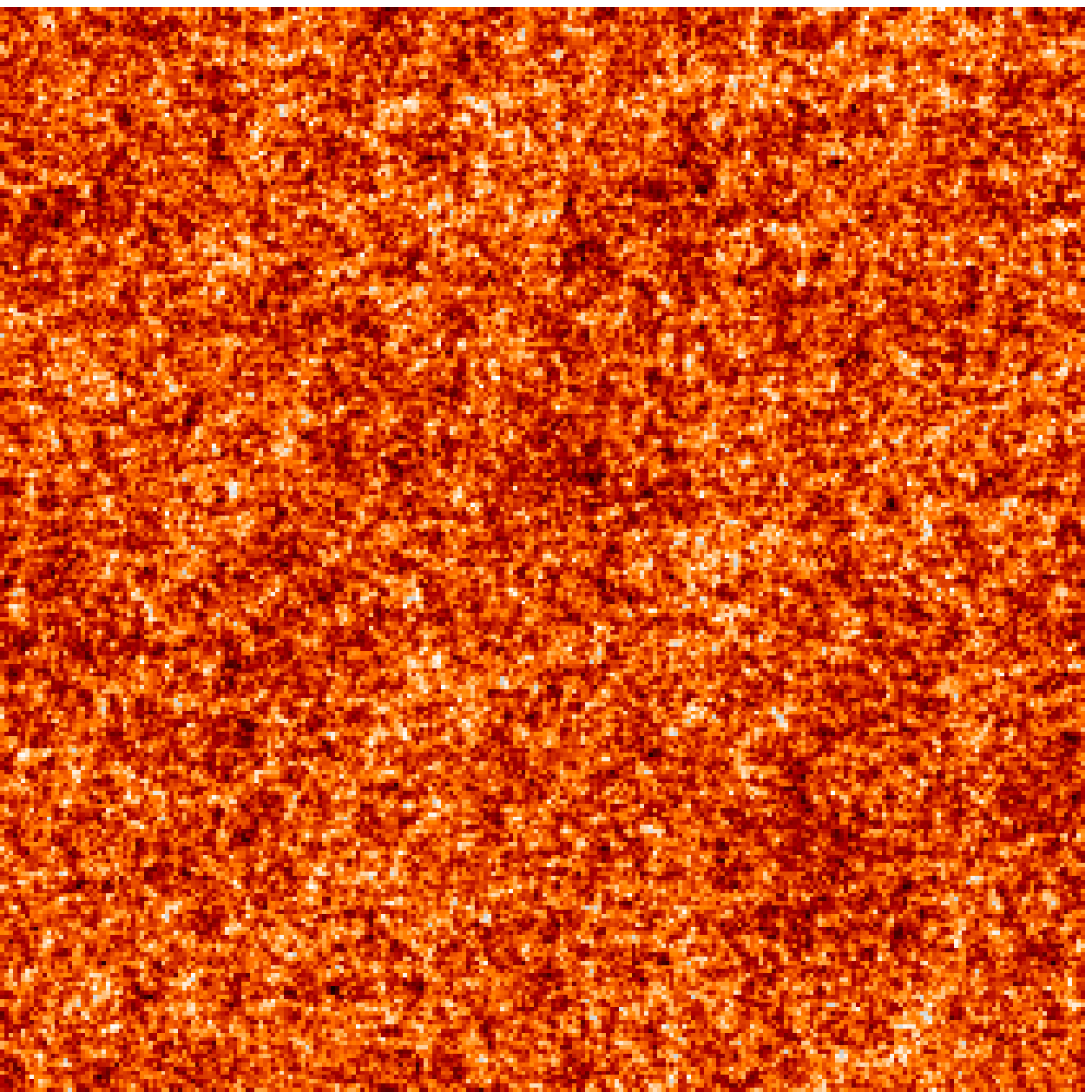}
\caption[]{Numerical simulation of galaxy clustering (left)
together with a version generated  randomly reshuffling the phases
between Fourier modes of the original picture (right).}
\label{eps1}
\end{figure}
Since the amplitude of each Fourier mode is unchanged in the phase
reshuffling operation, these two pictures have exactly the same
power-spectrum, $P(k)\propto|\tilde{\delta}({\bf k})|^2$. In fact,
they have more than that: they have exactly the same amplitudes
for all ${\bf k}$. They also have totally different morphology.
Further demonstrations of the importance of Fourier phases in
defining clustering morphology are given by Chiang (2001). The
evident shortcomings of $P(k)$ can be partly ameliorated by
defining higher-order quantities such as the
bispectrum~\cite{peeb,mvh,scf,vwhk} or correlations of
$\tilde{\delta}({\bf k})^2$ \cite{stir}.

\subsection{The Bispectrum and Phase Coupling} The bispectrum and
higher-order polyspectra vanish for Gaussian fields, but in a
non-Gaussian field they may be non-zero. The usefulness of these
and related quantities therefore lies in the fact that they encode
some information about non-linearity and non-Gaussianity.  To
understand the relationship between the bispectrum and Fourier
phases, it is very helpful to consider the following toy examples.
Imagine a simple density field defined in one spatial dimension
that consists of the superposition of two cosine components:
\begin{equation}
\delta(x) = A_1 \cos (\lambda_1x+\phi_1)+A_2 \cos (\lambda_2x +
\phi_2). \label{eq:toy1}
\end{equation}
The generalisation to several spatial dimensions is trivial. The
phases $\phi_1$ and $\phi_2$ are random and $A_1$ and $A_2$ are
constants. We can simplify the following by introducing a new
notation
\begin{equation} \delta(x)= A_1 \left(\begin{array}{c} \lambda_1
\\
\phi_1 \end{array} \right) + A_2 \left(\begin{array}{c} \lambda_2
\\ \phi_2 \end{array} \right). \label{eq:toy2} \end{equation}
Clearly this example displays no phase correlations. Now consider
a new field obtained from the example (\ref{eq:toy1}) through the
non-linear transformation
\begin{equation}
\delta(x)\mapsto \delta(x)+\epsilon \delta^2(x), \label{eq:nl}
\end{equation}
where $\epsilon$ is a constant parameter. Equation (\ref{eq:nl})
may be thought of as a very phenomenological representation of a
perturbation series, with $\epsilon$ controlling the level of
non-linearity. Using the same notation as equation
(\ref{eq:toy2}), the new field $\delta(x)$ can be written
\begin{eqnarray}
\delta(x) & = & B_1 \left(\begin{array}{c} \lambda_1 \\ \phi_1
\end{array} \right) + B_2 \left(\begin{array}{c} \lambda_2 \\
\phi_2 \end{array} \right) + B_3 \left(\begin{array}{c} 2\lambda_1
\\ 2\phi_1 \end{array} \right) + B_4 \left(\begin{array}{c}
2\lambda_2 \\ 2\phi_2 \end{array} \right) + \nonumber \\ & & + B_5
\left(\begin{array}{c} \lambda_1 + \lambda_2 \\ \phi_1 + \phi_2
\end{array} \right) + B_6 \left(\begin{array}{c}
\lambda_1-\lambda_2 \\ \phi_1-\phi_2 \end{array} \right),
\label{eq:quadro}
\end{eqnarray}
where the $B_i$ are constants obtained from the $A_i$. Notice in
equation (\ref{eq:quadro}) that the phases follow the same kind of
harmonic relationship as the wavenumbers. This form of phase
association is termed {\em quadratic} phase coupling. It is this
form of phase relationship that appears in the bispectrum. To see
this, consider another two toy examples. First, model A,
\begin{equation}
\delta_{\rm A}(x) = \left(\begin{array}{c} \lambda_1 \\ \phi_1
\end{array} \right) + \left(\begin{array}{c} \lambda_2 \\ \phi_2
\end{array} \right) + \left(\begin{array}{c} \lambda_3 \\ \phi_3
\end{array} \right), \label{eq:modelA}
\end{equation}
in which $\lambda_3=\lambda_1+\lambda_2$ but in which $\phi_1$,
$\phi_2$ and $\phi_3$ are random; and
\begin{equation}
\delta_{\rm B}(x) = \left(\begin{array}{c} \lambda_1 \\ \phi_1
\end{array} \right) + \left(\begin{array}{c} \lambda_2 \\ \phi_2
\end{array} \right) + \left(\begin{array}{c} \lambda_3=
\lambda_1+\lambda_2 \\ \phi_3= \phi_1 +\phi_2 \end{array} \right).
\label{eq:modelB}
\end{equation}
Model A exhibits no phase association; model B displays quadratic
phase coupling. It is straightforward to show that $\langle
\delta_A\rangle=\langle \delta_B\rangle=0$. The autocovariances
are equal:
\begin{equation}
\xi_A(r)=\langle\delta_A (x) \delta_A(x+r)\rangle =\xi_B(r) =
\frac{1}{2} [\cos (\lambda_1r) + \cos (\lambda_2r) + \cos
(\lambda_3r)],
\end{equation}
as are the power spectra, demonstrating that second-order
statistics are blind to phase association. The (reduced)
three-point autocovariance function is
\begin{equation}
\zeta(r_1, r_2)= \langle
\delta(x)\delta(x+r_1)\delta(x+r_2)\rangle.
\end{equation}
For model A we get
\begin{equation}
\zeta_A(r_1, r_2)=0,
\end{equation}
whereas for model B it is
\begin{eqnarray}
\zeta_B(r_1,r_2) & = &  \frac{1}{4} \left[ \cos(\lambda_2 r_1 +
\lambda_1 r_2) + \cos(\lambda_3 r_1 - \lambda_1 r_2) +
 \right.  \nonumber\\ & & +
\left. \cos(\lambda_1 r_1 + \lambda_2 r_2)  + \cos(\lambda_3 r_1 -
\lambda_2 r_2)  + \right.  \nonumber\\ & & + \left. \cos(\lambda_1
r_1 - \lambda_3 r_2) + \cos(\lambda_2 r_1 - \lambda_3 r_2)\right]
\end{eqnarray}

The bispectrum, $B(k_1,k_2)$, is defined as the two-dimensional
Fourier transform of $\zeta$, so $B_A(k_1,k_2)=0$ trivially,
whereas $B_B(k_1, k_2)$ consists of a single spike located
somewhere in the region of $(k_1, k_2)$ space defined by $k_2\geq
0$, $k_1\geq k_2$ and $k_1+k_2\leq \pi$. If $\lambda_1\geq
\lambda_2$ then the spike appears at $k_1=\lambda_1$,
$k_2=\lambda_2)$. Thus the bispectrum  measures the phase coupling
induced by quadratic nonlinearities. To reinstate the phase
information order-by-order requires an infinite hierarchy of
polyspectra.

An alternative way of looking at this issue is to note that the
information needed to fully specify a non-Gaussian field to
arbitrary order (or, in a wider context, the information needed to
define an image resides in the complete set of Fourier
phases~\cite{opp}. Unfortunately, relatively little is known about
the behaviour of Fourier phases in the nonlinear regime of
gravitational clustering~\cite{ryden,sms,soda,Jain,Jain2,cc2}, but
it is of great importance to understand phase correlations in
order to design efficient statistical tools for the analysis of
clustering data.

\subsection{Visualizing and Quantifying Phase Information} A vital
first step on the road to a useful quantitative description of
phase information is to represent it visually\cite{cc2}. In colour
image display devices, each pixel represents the intensity and
colour at that position in the image~\cite{thorn,fvd}. The
quantitative specification of colour involves three coordinates
describing the location of that pixel in an abstract colour space,
designed to reflect as accurately as possible the eye's response
to light of different wavelengths. In many devices this colour
space is defined in terms of the amount of Red, Green or Blue
required to construct the appropriate tone; hence the RGB colour
scheme. The scheme we are particularly interested in is based on
three different parameters: Hue, Saturation and Brightness. Hue is
the term used to distinguish between different basic colours
(blue, yellow, red and so on). Saturation refers to the purity of
the colour, defined by how much white is mixed with it. A
saturated red hue would be a very bright red, whereas a less
saturated red would be pink. Brightness indicates the overall
intensity of the pixel on a grey scale. The HSB colour model is
particularly useful because of the properties of the `hue'
parameter, which is defined as a circular variable. If the Fourier
transform of a density map has real part $R$ and imaginary part
$I$ then the phase for each wavenumber, given by
$\phi=\arctan(I/R)$, can be represented as a hue for that pixel
using the colour circle~\cite{cc2}.

The pattern of phase information revealed by this method related
to the gravitational dynamics of its origin. For example in our
analysis of phase coupling~\cite{cc} we introduced a quantity
$D_k$, defined by
\begin{equation}
D_k\equiv\phi_{k+1}-\phi_{k},
\end{equation}
which measures the difference in phase of modes with neighbouring
wavenumbers in one dimension. We refer to $D_k$ as the phase
gradient. To apply this idea to a two-dimensional simulation we
simply calculate gradients in the $x$ and $y$ directions
independently. Since the difference between two circular random
variables is itself a circular random variable, the distribution
of $D_k$ should initially be uniform. As the fluctuations evolve
waves begin to collapse, spawning higher-frequency modes in phase
with the original~\cite{sz}. These then interact with other waves
to produce a non-uniform distribution of $D_k$. For  examples, see
\begin{verbatim}
http://www.nottingham.ac.uk/~ppzpc/phases/index.html.
\end{verbatim}

It is necessary to develop quantitative measures of phase
information that can describe the structure displayed in the
colour representations. In the beginning the phases $\phi_k$ are
random and so are the $D_k$ obtained from them. This corresponds
to a state of minimal information, or in other words maximum
entropy. As information flows into the phases  the information
content must increase and the entropy decrease. One way to
quantify this is by defining an information entropy on the set of
phase gradients. One constructs a frequency distribution, $f(D)$
of the values of $D_k$ obtained from the whole map. The entropy is
then defined as
\begin{equation} S(D)=-\int f(D)\log [f(D)] dD,
\end{equation}
where the integral is taken over all values of $D$, i.e. from $0$
to $2\pi$. The use of $D$, rather than $\phi$ itself, to define
entropy is one way of accounting for the lack of translation
invariance of $\phi$, a problem that was missed in previous
attempts to quantify phase entropy~\cite{pm}. A uniform
distribution of $D$ is a state of maximum entropy (minimum
information), corresponding to Gaussian initial conditions (random
phases). This maximal value of $S_{\rm max}=\log(2\pi)$ is a
characteristic of Gaussian fields. As the system evolves it moves
into to states of greater information content (i.e. lower
entropy). The scaling of $S$ with clustering growth displays
interesting properties~\cite{cc}, establishing an important link
between the spatial pattern and the physics driving clustering
growth.

\section{Bias and Hierarchical Clustering}
The biggest stumbling-block for attempts to confront theories of
cosmological structure formation with observations of galaxy
clustering is the uncertain and possibly biased relationship
between galaxies and the distribution of gravitating matter. The
idea that galaxy formation might be biased goes back to the
realization by Kaiser (1984) that the reason Abell clusters
display stronger correlations than galaxies at a given separation
is that these objects are selected to be particularly dense
concentrations of matter. As such, they are very rare events,
occurring in the tail of the distribution function of density
fluctuations. Under such conditions a ``high-peak'' bias prevails:
rare high peaks are much more strongly clustered than more typical
fluctuations (Bardeen et al. 1986). If the properties of a galaxy
(its morphology, color, luminosity) are influenced by the density
of its parent halo, for example, then differently-selected
galaxies are expected to a different bias (e.g. Dekel \& Rees
1987). Observations show that different kinds of galaxy do cluster
in different ways (e.g. Loveday et al. 1995; Hermit et al. 1996).

In {\em local bias} models, the propensity of a galaxy to form at
a point where the total (local) density of matter is $\rho$ is
taken to be some function $f(\rho)$ (Coles 1993, hereafter C93;
Fry \& Gaztanaga 1993, hereafter FG93). It is possible to place
stringent constraints on the effect this kind of bias can have on
galaxy clustering statistics without making any particular
assumption about the form of $f$. In this {\it Letter}, we
describe  the results of a different approach to local bias models
that exploits new results from the theory of hierarchical
clustering in order to place stronger constraints on what a local
bias can do to galaxy clustering. We leave the technical details
to  Munshi et al. (1999a,b) and Bernardeau \& Schaeffer (1999);
here we shall simply motivate and present the results and explain
their importance in a wider context.

\subsection{Hierarchical Clustering} The fact that Newtonian
gravity is scale-free suggests that the $N$--point correlation
functions of self-gravitating particles, $\xi_N$, evolved into the
large-fluctuation regime by the action of gravity, should obey a
scaling relation of the form
\begin{equation}
\xi_p( \lambda {\bf r}_1, \dots  \lambda {\bf r}_p ) =
\lambda^{-\gamma(p-1)} \xi_p( {\bf r}_1, \dots {\bf r}_p )
\label{hierarchical}
\end{equation}
when the elements of a structure are scaled by a factor $\lambda$
(e.g. Balian \& Schaeffer 1989). Observations offer some support
for such an idea, in that the observed two-point correlation
function $\xi(r)$ of galaxies is reasonably well represented by a
power law over a large range of length scales,
\begin{equation} \xi({\bf r}) = \Big ( {r \over 5h^{-1} {\rm Mpc}}
\Big )^{-1.8} \end{equation} (Groth \& Peebles 1977; Davis \&
Peebles 1977) for $r$ between, say, $100 h^{-1} {\rm kpc}$ and $
10h^{-1}~{\rm~ Mpc}$. The observed three point function, $\xi_3$,
is well-established to have a hierarchical form \begin{equation}
\xi_3({\bf x}_a, {\bf x}_b, {\bf x}_c) = Q[\xi_{ab}\xi_{bc} +
\xi_{ac}\xi_{ab} + \xi_{ac}\xi_{bc}], \end{equation} where
$\xi_{ab}=\xi ({\bf x}_a, {\bf x}_b)$, etc, and $Q$ is a constant
(Davis \& Peebles 1977; Groth \& Peebles 1977). The four-point
correlation function can be expressed as a combination of  graphs
with two different topologies -- ``snake'' and ``star'' -- with
corresponding (constant) amplitudes $R_a$ and $R_b$ respectively:
\begin{eqnarray} \xi_4({\bf x}_a, {\bf x}_b, {\bf x}_c, {\bf x}_d) &
=& R_a[\xi_{ab}\xi_{bc}\xi_{cd} + \dots ({\rm
12~terms})]\nonumber\\ & &  + R_b[\xi_{ab}\xi_{ac}\xi_{ad} + \dots
({\rm 4~terms})]
\end{eqnarray} (e.g. Fry \& Peebles 1978; Fry 1984).

It is natural to guess that all p-point correlation functions can
be expressed as a sum over all possible p-tree graphs with (in
general) different amplitudes $Q_{p,\alpha}$  for each tree
diagram topology $\alpha$. If it is further assumed that there is
no  dependence of these amplitudes upon the shape of the diagram,
rather than its topology, the correlation functions should obey
the following relation: \begin{equation} \xi_p( {\bf r}_1, \dots
{\bf r}_p ) = \sum_{\alpha, ~p-{\rm trees}} Q_{p,\alpha} \sum_{\rm
labellings} \prod_{\rm edges}^{(p-1)} \xi({\bf r}_i, {\bf r}_j).
\end{equation} To go further it is necessary to find a way of
calculating $Q_p$. One possibility, which appears remarkably
successful when compared with numerical experiments (Munshi et al.
1999b; Bernardeau \& Schaeffer 1999), is to calculate the
amplitude for a given graph by simply assigning a weight to each
vertex of the diagram $\nu_n$, where $n$ is the order of the
vertex (the number of lines that come out of it), regardless of
the topology of the diagram in which it occurs. In this case
\begin{equation} Q_{p,\alpha}=\prod_{\rm vertices} \nu_n.
\end{equation} Averages of higher-order correlation functions can be
defined as \begin{equation} \bar{\xi}_p=\frac{1}{V^p} \int \ldots
\int \xi_p({\bf r}_1\ldots {\bf r_p}) dV_1\ldots dV_p.
\label{eq:xibar}
\end{equation} Higher-order statistical properties of galaxy counts
are often described in terms of the scaling parameters $S_p$
constructed from the $\bar{\xi}_p$ via \begin{equation}
S_p=\frac{\bar{\xi}_p}{\bar{\xi}_2^{p-1}}.\label{eq:s_p}
\end{equation} It is a consequence of the particular class of hierarchical clustering models
defined by equations (5) \& (6) that {\it all} the $S_p$ should be
constant, independent of scale.

\subsection{Local Bias}

Using a generating function technique \cite{bs92} it is possible
to derive a series expansion for the $m$-point count probability
distribution function of the objects $P_m(N_1,....N_m)$ (the joint
probability of finding $N_i$ objects in the $i$-th cell, where $i$
runs from $1$ to $m$) from the $\nu_n$. The hierarchical model
outlined above is therefore statistically complete. In principle,
therefore, any statistical property of the evolved distribution of
matter can be calculated just as it can for a Gaussian random
field. This allows us to extend various results concerning the
effects of biasing on the initial conditions into the nonlinear
regime in a more elegant way than is possible using other
approaches to hierarchical clustering.

For example, let us consider the joint  probability $P_2(N_1,
N_2)$ for two cells to contain $N_1$ and $N_2$ particles
respectively. Using the generating-function approach outlined
above, it is quite easy to show that, at lowest order,
\begin{equation}
P_2(N_1,N_2)=P_1(N_1)P_1(N_2)+P_1(N_1)b(N_1)P_1(N_2)b(N_2)\xi_{12}(r_{12
}), \label{eq:2pt} \end{equation}
 where the $P_1(N_i)$ are the
individual count probabilities of each volume separately and
$\xi_{12}$ is the underlying mass correlation function. The
function $b(N_i)$ we have introduced in (9) depends on the set of
$\nu_n$ appearing in equation (6); its precise form does not
matter in this context, but the structure of equation (9) is very
useful. We can use (9) to define
\begin{equation}
1+\xi_{N_1 N_2}(r_{12}) \equiv \frac{P(N_1,
N_2)}{P_1(N_1)P_1(N_2)},
\end{equation}
where $\xi_{N_1 N_2}(r_{12})$ is the cross-correlation of
``cells'' of occupancy $N_1$ and $N_2$ respectively. From this
definition and equation (9) it follows that
\begin{equation} \xi_{N_1 N_2}(r)=b(N_1)b(N_2)\xi_{12}(r);
\label{eq:2bias}
\end{equation}
we have dropped the subscripts on $r$ for clarity from now on.
>From (11) we can obtain
\begin{equation}
b_{N}^2(r_{12}) = \frac{\xi_{N N}(r)}{\xi_{12}(r)}
\end{equation}
for the special case where $N_1=N_2=N$ which can be identified
with the usual definition of the bias parameter associated with
the correlations among a given set of objects $\xi_{\rm
obj}(r)=b^{2}_{\rm obj} \xi_{\rm mass}(r)$. Moreover, note that at
this order (which is valid on large scales), the correlation bias
defined by equation (11) factorizes into contributions $b_{N_i}$
from each individual cell (Bernardeau 1996; Munshi et al. 1999b).

Coles (1993) proved, under weak conditions on the form of a local
bias $f(\rho)$ as discussed in the introduction, that the
large-scale biased correlation function would generally have a
leading order term proportional to $\xi_{12}(r_{12})$. In other
words, one cannot change the large-scale slope of the correlation
function of locally-biased galaxies with respect to that of the
mass. This ``theorem'' was proved for bias applied to Gaussian
fluctuations only and therefore does not obviously apply to galaxy
clustering, since even on large scales deviations from Gaussian
behaviour are significant. It also has a more minor loophole,
which is that for certain peculiar forms of $f$ the leading order
term is proportional to $\xi_{12}^2$, which falls off more sharply
than $\xi_{12}$ on large scales.

Steps towards the plugging of this gap began with FG93 who used an
expansion of $f$ in powers of $\delta$ and weakly non-linear
(perturbative) calculations of $\xi_{12}(r)$ to explore the
statistical consequences of biasing in more realistic (i.e.
non-Gaussian) fields. Based largely on these  arguments, Scherrer
\& Weinberg (1998), hereafter SW98, confirmed the validity of the
C93 result in the non-linear regime, and also showed explicitly
that non-linear evolution always guarantees the existence of a
linear leading-order term regardless of $f$, thus plugging the
small gap in the original C93 argument. These works have a similar
motivation the approach I am discussing here, and also exploit
hierarchical scaling arguments of the type discussed above {\em en
route} to their conclusions. What is different about the approach
we have used in this paper is that the somewhat cumbersome
simultaneous expansion of $f$ and $\xi_{12}$ used by SW98 is not
required in this calculation. The generating functions to proceed
directly to the joint probability (9), while SW98 have to perform
a complicated sum over moments of a bivariate distribution. The
factorization of the probability distribution (9) is also a
stronger result than that presented by SW98, in that it leads
almost trivially to the C93 ``theorem'' but also generalizes to
higher-order correlations than the two-point case under discussion
here.

Note that the density of a cell of given volume is simply
proportional to its occupation number $N$. The factorizability of
the dependence of $\xi_{N_1 N_2}(r_{12})$ upon $b(N_1)$ and
$b(N_2)$ in (11) means that applying a local bias $f(\rho)$ boils
down to applying some bias function $F(N)=f[b(N)]$ to each cell.
Integrating over all $N$ thus leads directly to the same
conclusion as C93, i.e. that the large-scale $\xi(r)$ of
locally-biased objects is proportional to the underlying matter
correlation function. This has also been confirmed by numerically
using $N$-body experiments (Mann et al. 1998; Narayanan et al.
1998).

\subsection{Halo Bias}

In hierarchical models, galaxy formation involves the following
three stages:
\begin{enumerate}
\item the formation of a dark matter halo;
\item the settling of gas into the halo potential;
\item the cooling and fragmentation of this gas into stars.
\end{enumerate}
Rather than attempting to model these stages in one go by a simple
function $f$ of the underlying density field it is interesting to
see how each of these selections might influence the resulting
statistical properties. Bardeen et al. (1986), inspired by Kaiser
(1984), pioneered this approach by calculating detailed
statistical properties of high-density regions in Gaussian
fluctuations fields. Mo \& White (1996) and Mo et al. (1997) went
further along this road by using an extension of the
Press-Schechter (1974) theory to calculate the correlation bias of
halos, thus making an attempt to correct for the dynamical
evolution absent in the Bardeen et al. approach. The extended
Press-Schechter approach seem to be in good agreement with
numerical simulations, except for small halo masses (Jing 1998).
It forms the basis of many models for halo bias in the subsequent
literature (e.g. Moscardini et al. 1998; Tegmark \& Peebles 1998).

The hierarchical models furnish an elegant extension of this work
that incorporates both density-selection and non-linear dynamics
in an alternative to the  Mo \& White (1996) approach. We  exploit
the properties of equation (\ref{eq:2pt}) to construct the
correlation function of volumes where the occupation number
exceeds some critical value. For very high occupations these
volumes should be in good correspondence with collapsed objects.

The way of proceeding is to construct a tree graph for all the
points in both volumes. One then has to re-partition the elements
of this  graph into internal lines (representing the correlations
within each cell) and external lines (representing inter-cell
correlations). Using this approach the distribution of
high-density regions in a field whose correlations are given by
eq. (5) can be shown to be itself described by a hierarchical
model, but one in which the vertex weights, say $M_n$, are
different from the underlying weights $\nu_n$ (Bernardeau \&
Schaeffer 1992, 1999; Munshi et al. 1999a,b).

First note that a density threshold is in fact a form of local
bias, so the effects of halo bias are governed by the same
strictures as described in the previous section. Many of the other
statistical properties of the distribution of dense regions can be
reduced to a dependence on a scaling parameter $x$, where
\begin{equation} x=N/N_c. \end{equation} In this definition
$N_c=\bar{N}\bar{\xi}_2$, where $\bar{N}$ is the mean number of
objects in the cell and $\bar{\xi}_2$ is defined by eq.
(\ref{eq:xibar}) with $p=2$.  The scaling parameters $S_p$  can be
calculated as functions of $x$, but are generally rather messy
(Munshi et al. 1999a). The most interesting limit when $x\gg 1$
is, however, rather simple. This is because the vertex weights
describing the distribution of halos depend only on the $\nu_n$
and this dependence cancels in the ratio (\ref{eq:s_p}). In this
regime, \begin{equation} S_p=p^{(p-2)} \end{equation} for all
possible hierarchical models. The reader is referred to Munshi et
al. (1999a) for details. This result is also obtained in the
corresponding limit for very massive halos by Mo et al. (1997).
The agreement between these two very different calculations
supports the inference that this is a robust prediction for the
bias inherent in dense regions of a distribution of objects
undergoing gravity-driven hierarchical clustering.

\subsection{Progress on Biasing}

The main purpose of this lecture has been to discuss recent
developments in the theory of gravitational-driven hierarchical
clustering. The model described in equations (5) \& (6) provides a
statistically-complete prescription for a density field that has
undergone hierarchical clustering. This allows us to improve
considerably upon biasing arguments based on an underlying
Gaussian field.

These methods allow a simpler proof of the result obtained by SW98
that  strong non-linear evolution does not invalidate the local
bias theorem of C93. They also imply that the effect of bias on a
hierarchical density field is factorizable. A special case of this
is the bias induced by selecting regions above a density
threshold. The separability of bias predicted in this kind of
model could be put to the test if a population of objects could be
found whose observed characteristics (luminosity, morphology,
etc.) were known to be in one-to-one correspondence with the halo
mass. Likewise, the generic prediction of higher-order correlation
behaviour described by the behaviour of $S_p$ in equation
(\ref{eq:s_p}) can also be used to construct a test of this
particular form of bias.

Referring to the three stages of galaxy formation described in \S
4, analytic theory has now developed to the point where it is
fairly convincing on (1) the formation of halos. Numerical
experiments are beginning now to handle (2) the behaviour of the
gas component (Blanton et al. 1998, 1999). But it is unlikely that
much will be learned about (3) by theoretical arguments in the
near future as the physics involved is poorly understood (though
see Benson et al. 1999). Arguments have already been advanced to
suggest that bias might not be a deterministic function of $\rho$,
perhaps because of stochastic or other hidden effects (Dekel \&
Lahav 1998; Tegmark \& Bromley 1999). It also remains possible
that large-scale non-local bias might be induced by environmental
effects (Babul \& White 1991; Bower et al. 1993).

Before adopting these more complex models, however, it is
important to exclude the simplest ones, or at least deal with that
part of the bias that is attributable to known physics. At this
stage this means that the `minimal' bias model should be that
based on the selection of dark matter halos. Establishing the
extent to which observed galaxy biases can be explained in this
minimal way is clearly an important task.

\section{Discussion}

In fairly recent history,  cosmological data sets were sparse and
incomplete, and the statistical methods deployed to analyse them
were crude. Second-order statistics, such as $P(k)$ and $\xi(r)$,
are blunt instruments that throw away the fine details of the
delicate pattern of cosmic structure. These details lie in the
distribution of Fourier phases to which second-order statistics
are blind. It would not do justice to massively improved data if
effort were directed only to better estimates of these quantities.
Moreover, as we have shown, phase information provides a unique
fingerprint of gravitational instability developed from Gaussian
initial conditions (which have maximal phase entropy). Methods
such as those  described above can therefore be used to test this
standard paradigm for structure formation. They can also furnish
direct tests of the presence of initial non-Gaussianity
~\cite{fmg,pvf,bt}.

But there is also an important general point to be made about the
philosophy of large-scale structure studies. The existing
approaches are dominated by a {\em direct} methodology. A
hypothetical mixture of ingredients is constructed (see Section
2.6), and {\em ab initio} simulations used to propagate the
initial conditions to a model of reality that would pertain if the
model were true. If it fails, one revises the model. But there are
now many models which agree more-or-less with the existing data.
These also contain free parameters that can be used to massage
them into compliance with observations. In particular, we can
appeal to a complex non-linear and non-local bias to achieve this.
The usefulness of these direct hypothesis tests is therefore open
to doubt.

The stumbling block lies with the fact that we still cannot
reliably predict the relationship between galaxies and mass.
Although theory seems to have slowed down, we do now have the
prospect of huge amounts of data arriving on the scene. A better
approach than the direct one I have mentioned is to treat those
unknown aspects of galaxy formation as an inverse problem. Given a
sufficiently flexible and realistic model we should infer
parameter values from observations. To exploit this approach
requires the development of simple models that can be used to
close the inductive loop connecting theory with observations. For
this reason it is important to continue constructing simple models
of bias and galaxy clustering generally, since these are such
valuable inferential tools.

As the raw material is increasing in both quality and quantity, it
is time to refine our statistical technology so that the subtle
and precious artifacts previously ignored can be both detected and
extracted.

\end{article}
\end{document}